\documentclass[11pt]{article}

\usepackage[accepted]{tmlr}

\usepackage{mdframed}
\usepackage{times}
\usepackage{lipsum}
\usepackage{setspace}
\usepackage{geometry}
\usepackage{graphicx}
\usepackage{url} 
\usepackage{longtable}
\usepackage{amsmath}
\usepackage{booktabs}
\usepackage{amsfonts}
\usepackage{amssymb}
\usepackage{amsmath}
\usepackage{dsfont}
\usepackage{mathrsfs}  
\usepackage{xcolor,soul}

\newtheorem{example}{Example}
\newtheorem{fallacy}{Fallacy}
\usepackage{hyperref}
\hypersetup{
  colorlinks=true,   
  urlcolor=blue,     
  linkcolor=black,   
  citecolor=black    
}

\newmdenv[
  leftline=true,
  topline=false,
  bottomline=false,
  rightline=false,
  linewidth=2pt,
  linecolor=gray!50,
  innerleftmargin=10pt,
  innertopmargin=0pt,
  innerbottommargin=0pt,
  skipabove=6pt
]{fallacynote}

\title{Why Customer Choice Models Matter}

\author{\name Berry Gerrits 
\email b.gerrits@utwente.nl \\
      \addr Industrial Engineering and Management Science\\
      University of Twente
\AND 
\name Fabian Akkerman
\email f.r.akkerman@utwente.nl \\
      \addr Industrial Engineering and Management Science\\
      University of Twente}

\begin{document}
\maketitle

\begin{abstract}
Customer choice models are central to revenue management in attended home delivery, yet the field usually picks one without much thought. Does the choice model matter? We believe it does, and this paper shows that revenue management policies judged by their own assumed model, or trusted under a single set of parameters, can be worse than they look.
\end{abstract}

\section{Introduction}
A Customer Choice Model (CCM) determines how a human selects a choice (or series of choices) among a set of options \cite[p.~3]{Train2009}. For example, a consumer chooses which of a finite list of delivery time slots to select for grocery delivery in Attended Home Delivery (AHD). CCMs are foundational to many of the Revenue Management (RM) models used in AHD. As a consequence, the assumed CCM can heavily determine the outcome of any method used. Most, if not all, influential AHD works (e.g., \citet{campbell2006}, \citet{agatz2011}, \citet{ehmke2014}, \citet{yang2016, yang2017}, \citet{klein2019}) seem to treat the CCM as a necessary modelling choice that does not deserve much thought, and focus on pricing policies, slotting policies, and cost approximation methods instead. 

In this paper, we argue that the choice of CCM may severely impact the validity of the outcomes of the proposed algorithmic machinery, potentially undermining the validity of the resulting insights. This hypothesis originates from the fact that although a CCM is always a simplification of reality (as one only observes noisy observations from true human behaviour), using oversimplified models may depart significantly from reality. Moreover, in many AHD works, authors assume the CCM parameters to be visible to the RM policy. These policies are subsequently evaluated with the same CCM, even though the true (unknown) CCM may significantly differ. To make matters worse, sensitivity analyses on the CCM remain strikingly absent in the literature. Consequently, claims that the proposed pricing or slotting policies are \emph{(close-to) optimal} may be overconfident. 

In this paper, we establish three fallacies (i) assuming perfect customer choice knowledge (i.e., the CCM used to derive policies is the same as the CCM used to evaluate these policies) can result in wrong insights, (ii) even when using the perfect model, uncertainty in the model parameters can still lead to misleading results, and (iii) modelling customer segments provides better granularity to CCMs. In order to do so, we first establish a common view on the relevant aspects of the decision-making process in AHD. 

\subsection{A Unified View of Attended Home Delivery}

In AHD, a company provides services or goods to customers at a specific (home) location during a specific delivery time slot in which they are available to receive the services or goods (e.g., groceries ordered through an e-commerce platform). The distinctive feature of AHD compared to other last-mile delivery services is the requirement of physical presence at the delivery location during the time slot. From the perspective of the customer, the fulfilment process goes as follows. First, the customer fills its online basket with groceries and selects the delivery location. Next, the customer is presented time slots for the delivery, typically over multiple days and each slot is associated with a price. Finally, the customer chooses a slot, completes the order, and is delivered within the selected time slot. 

From a business perspective, whenever a customer arrives, the retailer has to decide which time slots to offer the customer against which prices. This can be done in a \textit{static} fashion in which all decisions are made before any customer has arrived, or \textit{dynamically} in which the offered time slots and/or the associated prices may be updated after every booking. This booking period is characterized by a \textit{cut-off time} after which customers are no longer allowed to select time slots for a delivery day. After the cut-off time, the company decides upon the delivery route by solving a Vehicle Routing Problem with Time Windows (VRPTW) as introduced by \cite{Solomon1987}, where the time windows coincide with the time slots selected by the customers. Due to the fact that customer demand is not known a priori, the retailer needs to make decisions regarding time slot offering, pricing, and possible routing policies under uncertainty. 


It may come as no surprise that the main goal of a retailer offering AHD is to maximize its profit through some form of policy optimization. Although many methods exists in the literature, differing considerably in their algorithms and assumptions, we argue that the most commonly used methods for AHD can be viewed as instances of a common optimization framework, consisting of four components: 

\begin{enumerate}
\itemsep0em 
    \item a \textbf{slotting policy} that determines which time slots are offered to an arriving customer;
    \item a \textbf{pricing policy} that determines the delivery fee associated with each offered slot;
    \item a \textbf{customer choice model} that models the customers' response to the offering;
    \item a \textbf{cost approximation model} that estimates the expected fulfilment cost associated with the order.
\end{enumerate}

The first two components represent the retailer's main decision variables, while the latter two provide the foundation for these decisions. To formulate this four-component revenue management problem as a generic optimization problem (i.e., maximize profits), we require some mathematical notation. Let $S$ denote the set of feasible delivery time slots and $\mathbf{x}=(x_s)_{s\in S}$ denote the vector of slot availability decisions, where each component corresponds to one delivery slot:

\begin{equation*}
        x_s=
        \begin{cases}
        1,&\text{if slot } s \text{ is offered},\\
        0,&\text{otherwise}.
        \end{cases}
\end{equation*}

Let $\mathbf{p}=(p_s)_{s\in S}$ denote the vector of delivery prices, $\pi_s(\mathbf{x},\mathbf{p})$ denote the probability that the customer selects slot $s$ under the offered assortment and prices, and $\pi_0(\mathbf{x},\mathbf{p})$ denote the probability that no purchase is made. Finally, let ${\mathcal{C}}(\mathbf{x},\mathbf{p})$ denote the expected fulfilment cost resulting from the accepted order.
After observing that the customer choice probabilities satisfy 
$\pi_0(\mathbf{x},\mathbf{p})+\sum_{s\in S}\pi_s(\mathbf{x},\mathbf{p})=1$, the expected revenue generated by a customer arrival equals:
\begin{equation*}
\mathcal{R}(\mathbf{x},\mathbf{p})=\sum_{s\in S}\pi_s(\mathbf{x},\mathbf{p})\,p_s,
\end{equation*}
and through the miracle of basic calculus, the expected profit is obtained by:
\begin{equation*}
\mathscr{P} = \sum_{s\in S}\pi_s(\mathbf{x},\mathbf{p})\,p_s -  \mathcal{C}(\mathbf{x},\mathbf{p}).
\end{equation*}
The revenue management problem can therefore be formulated as:
\begin{align*}
\max_{\mathbf{x},\mathbf{p}}\quad
&
\sum_{s\in S}
\pi_s(\mathbf{x},\mathbf{p})\,p_s
-
\mathcal{C}(\mathbf{x},\mathbf{p}),
\label{eq:genericRM}
\\
\text{s.t.}\quad
&
\mathbf{x} \in\mathcal X,
\\
&
\mathbf{p}\in\mathcal P,
\end{align*}
where $\mathcal X$ denotes the feasible set of slot offerings and $\mathcal P$ the feasible pricing decisions. The customer choice model and the cost approximation are embedded in the functions $\pi(\cdot)$ and $\mathcal C(\cdot)$, respectively.

This optimization model remains unchanged across virtually all AHD revenue management models. Different methods correspond to different choices for each of the four components. The validity of this claim may demand some decryption, which we offer below. 

Recall that the AHD literature resembles around two main components: a slotting policy and a pricing policy, both of which may be static or dynamic, resulting in four distinct combinations. As a matter of fact, AHD literature is often positioned based on these four combinations, as shown in Table~\ref{tab:agatz_framework}. Also recall that the customer choice model and the cost approximation model underlie the slotting and pricing policies.  

\begin{table}[!h]
    \centering
    \caption{Constraint types yielding the four common AHD methods.}
    \begin{tabular}{c|cc}
     & Static pricing & Dynamic pricing\\
    \hline
    Static slotting &
    $\mathbf{x}=\mathbf{x}^{\mathrm{stat}}$ &
    $\mathbf{x}=\mathbf{x}^{\mathrm{stat}},\;\mathbf{p}=\mathbf{p}(\mathbf{z})$\\
    Dynamic slotting &
    $\mathbf{x}=\mathbf{x}^{\mathrm{stat}},\;\mathbf{p}=\mathbf{p}^{{stat}}$ &
    $\mathbf{x}=\mathbf{x}^{\mathrm{stat}},\;\mathbf{p}=\mathbf{p}(\mathbf{z})$
    \end{tabular}
    \label{tab:agatz_framework}
\end{table}

First, the slotting policy determines the availability vector $\mathbf{x}$. In the case of static slotting, the slot availability is determined offline before the ordering period start, resulting in $\mathbf{x}=\mathbf{x}^{\mathrm{stat}}$, and remains fixed for all customers. In the case of dynamic slotting, the offered slots depend on the current booking state $\mathbf{z}$ (e.g., accepted orders, remaining capacity, or partial route construction), thus $\mathbf{x}=\mathbf{x}(\mathbf{z})$ allows time slot availability to adapt through the ordering period. 

Second, the pricing policy determines the delivery fees. In the case of static pricing, we obtain $\mathbf{p}=\mathbf{p}^{{stat}}$. Note that \textit{static} does not infer that each time slot $s$ has the same price, it merely states that the prices are fixed before any booking occurs. Alternatively, in the case of dynamic pricing, prices depend on the current system state $\mathbf{z}$: $\mathbf{p}=\mathbf{p}(\mathbf{z})$. 


The above framework is commonly used in AHD literature to position a work in terms of the pricing and slotting policy under consideration, recently also integrating the operational, tactical, and strategic planning levels \citep{wamuth2023}. Notably, the underlying CCM attracts little attention, even though it is an important input to both policy types. As stated above, in this paper we aim to disentangle the role CCMs play in AHD. The remainder of this paper is structured as follows: in Section~\ref{literature} we discuss AHD literature with a focus on CCMs, Section~\ref{experiemtns} shows numerical results, and Section~\ref{conclusion} concludes the paper.

\section{Literature Overview}\label{literature}
Having established that the customer choice model enters the revenue management problem through $\pi_s(\mathbf{x},\mathbf{p})$, we review the AHD literature specifically aimed at this component. We treat a customer $i$ as choosing a time slot $s$ from the offered set on the basis of a random utility $U_{is} = V_{is} + \varepsilon_{is}$, where $V_{is}$ is a deterministic component
depending on attributes of the customer--slot pair and $\varepsilon_{is}$ a residual capturing unobserved factors. The main models used in AHD are ank-based, conditional logit, multinomial logit (MNL), nested MNL, mixed logit, the basic attraction model (BAM), and the generalized attraction model (GAM). These differ only in how they specify $V_{is}$ and $\varepsilon_{is}$. We review each in turn together with its behaviour on the following running example, and then show they are all instances of a single framework, summarized in Table~\ref{tab:choice_model_positioning}. In our review, we evaluate which models are used in the recent and not-so-recent AHD literature. For brevity of exposition, we omit the index denoting the customer segment. 

\begin{example}
An arriving customer is offered a subset of three non-overlapping one-hour slots on a single delivery day, $S=\{A,B,C\}$ with $A$ (08:00--09:00), $B$ (12:00--13:00), and $C$ (17:00--18:00), alongside the no-purchase option indexed by $0$. The retailer prefers to fill slot $B$, as it is inserted cheaply into the current route, and offers a discount on $B$ by lowering $p_B$. We examine how each model maps the availability vector $\mathbf{x}$ and the price vector $\mathbf{p}$ into the probabilities $\pi_s(\mathbf{x},\mathbf{p})$ and $\pi_0(\mathbf{x},\mathbf{p})$.
\end{example}

\paragraph{Rank-based models}
\label{subsec:rank}
Rank-based models represent each customer by a deterministic preference list over all slot-price combinations, obtained by $\varepsilon_{is}\to 0$. Given the offered assortment, the customer selects the highest-ranked available alternative and refrains from purchasing only if none of their acceptable options is offered. In terms of our notation, the probabilities $\pi_s(\mathbf{x},\mathbf{p})$ are degenerate: for a given segment they place all mass on a single slot whenever that slot is available under $\mathbf{x}$, and on the no-purchase option otherwise. Customers are segmented by identical preference lists, which are assumed known a priori. This is for example used by \cite{klein2019} who adopt a general non-parametric rank-based model, segmenting customers by identical preference orderings over all combinations of time slots and price options. \cite{cleophas2014} adopt a static customer model in which each customer requests a single combination of delivery area and time slot; if this combination is not available, the customer abandons the order. Customer behaviour is thus modelled without choice flexibility, i.e., there is no option for the customer to select alternative time slots or respond to pricing incentives. Relatedly, \cite{ehmke2014} model choice only implicitly through a primary slot and a single fallback alternative from a fixed set, focusing on the provider-side feasibility of accommodating a request rather than on the probabilistic structure of $\pi_s(\cdot)$.
 
For the running example, a customer (segment) with preference list $[B \succ A \succ 0]$ books $B$ (deterministically) whenever $x_B=1$, irrespective of the discount, which enters only insofar as it shaped the ranking offline. If the retailer offers only $C$ (i.e., $x_C=1$, $x_A=x_B=0$), this customer does not book and all probability mass goes to $\pi_0$.

\paragraph{Multinomial logit (MNL)}
\label{subsec:mnl}
The multinomial logit model assumes i.i.d.\ Gumbel residuals, so that the utility of slot $s$ decomposes into a deterministic component $v_s$ capturing slot attributes and a Gumbel error, and the choice probabilities take the closed form

\begin{equation*}
\pi_s(\mathbf{x},\mathbf{p})
=
\frac{x_s\,e^{v_s(p_s)}}
     {e^{v_0}+\sum_{s \in S} x_s\,e^{v_s(p_s)}},
\qquad
\pi_0(\mathbf{x},\mathbf{p})
=
\frac{e^{v_0}}
     {e^{v_0}+\sum_{s \in S} x_s\,e^{v_s(p_s)}},
\end{equation*}
where $v_0$ denotes the utility of no purchase. Behaviour is typically assumed homogeneous within a customer segment. The model's principal limitation is the independence of irrelevant alternatives (IIA) property, under which withdrawing a slot from $\mathbf{x}$ redistributes its probability across the remaining alternatives in fixed proportion, so that adding a 17:30 slot to a set
already containing a 17:00 slot draws demand proportionally from all slots rather than mostly from the similar 17:00 one, causing MNL to over-capture demand, especially when the assortment is narrow. IIA is a property many authors cite as a limitation and then assume anyway. Nonetheless, MNL is the proverbial workhorse of the AHD literature. \cite{yang2016} develop a dynamic slot-pricing strategy using a continuous MNL model, segmenting customers by their chosen delivery day and assuming homogeneous behaviour within a segment for reasons of tractability and price discrimination. In a follow-up, \cite{yang2017} retain the same MNL specification, post-processed to admit only non-overlapping slots, and contribute an approximate dynamic programming scheme for the opportunity cost embedded in $\mathcal{C}(\mathbf{x},\mathbf{p})$ rather than a richer choice model. \cite{asdemir2009} embed an MNL choice model within a Markov decision process for dynamic pricing, segmenting customers into a small number of classes characterized by delivery-day preference and price sensitivity.

In the running example, lowering $p_B$ raises $v_B$ and hence $\pi_B$. Because the denominator is common to all alternatives, the increase in $\pi_B$ impacts every other option proportionally: $\pi_A$, $\pi_C$, and $\pi_0$ each shrink by the same factor, illustrating the IIA property. 

\paragraph{Nested MNL}
\label{subsec:nested}
The nested logit model replaces the i.i.d.\ Gumbel residuals with a generalized
extreme value distribution that permits correlation within nests, partitioning the
alternatives into groups of correlated options and modelling choice as a sequential
decision. First the customer chooses between nests, then within the selected nest. \cite{strauss2020} model AHD customer behaviour through a nested logit structure, representing the delivery-slot decision as a sequential choice between premium and standard time windows, and construct dynamic pricing policies upon the estimated demand model.

Suppose the running example partitions $S$ into a standard nest $\{A,C\}$ and a flexible-discounted nest $\{B\}$. Reducing $p_B$ raises the inclusive value of the flexible nest and hence $\pi_B$. Crucially, withdrawing $A$ shifts probability mainly to the alternative in the $C$ nest, rather than uniformly to $B$ and $0$, distinguishing the nested model from plain MNL.

\paragraph{Mixed logit}
\label{subsec:mixed}
Mixed logit lets the coefficients vary across customers,
$\boldsymbol{\beta}_i \sim f(\cdot\mid\boldsymbol{\theta})$, so that the selection
probabilities are integrals of the logit expression over that distribution:
\begin{equation*}
\pi_s(\mathbf{x},\mathbf{p})
=
\int
\frac{x_s\,e^{v_s(p_s;\,\boldsymbol{\beta})}}
     {e^{v_0}+\sum_{s\in S} x_s\,e^{v_s(p_s;\,\boldsymbol{\beta})}}
\,
f(\boldsymbol{\beta})\,\mathrm{d}\boldsymbol{\beta}.
\end{equation*}
This captures unobserved heterogeneity and relaxes IIA at the population level, at the cost of simulation-based evaluation of $\pi_s(\cdot)$ whenever a closed form of $f(\boldsymbol{\beta})$ is absent or intractable. \cite{abdolhamidi2024mixed} introduce a tactical slot-assortment and price-discount problem for subscription-based e-retailers under mixed logit demand, integrating the assortment decision $\mathbf{x}$, the pricing decision $\mathbf{p}$, and routing within a simulation-based mixed-integer linear program. To contend with the simulation-based choice probabilities they develop an adaptive large neighbourhood search aligned with a sample average approximation reformulation.

In the running example, the customer population is a distribution rather than a single point. Some draws value to the early slot $A$, others to the discounted
midday slot $B$. Lowering $p_B$ captures the price-sensitive tail strongly while barely moving customers whose timing preference dominates, yielding a differentiated response that a single MNL cannot reproduce.

\paragraph{Conditional logit}
\label{subsec:conditional}
The conditional logit model shares the i.i.d.\ Gumbel residuals and softmax form of
MNL but describes each slot $s$ through an attribute vector $\mathbf{a}_s$ (e.g., in the AHD setting its timing, slot width, and price $p_s$) using $v_s=\boldsymbol{\beta}^{\top}\mathbf{a}_s$ with the coefficient vector
$\boldsymbol{\beta}$ common to all slots.
The utility thus only depends on attributes of the time slots rather than on the customer. \cite{amorim2024} estimate a conditional logit model on a large order set across eight delivery areas, characterizing each slot by speed (order-to-delivery time), precision (slot width), and timing (time of day and day of week).

In the running example, the model does not treat $A$, $B$, and $C$ as three unrelated options but describes each by the same attributes, e.g., timing, slot width, and price. Because price and timing share one utility scale, the estimated coefficients express the discount on $B$ as an equivalent amount of waiting: the reduction in $p_B$ is worth, say, a given number of hours of less convenient timing. The output is therefore a willingness-to-wait that transfers across slots and customers, rather than a fixed preference constant attached to each slot.

\paragraph{Basic attraction model (BAM)}
\label{subsec:bam}
Thus far we have presented our discussion from a random-utility perspective. As an alternative, \emph{attraction models} are also used in the AHD literature. Rather than deriving choice from utilities and error terms, an attraction model assigns to each alternative a non-negative attraction value and sets the choice probability equal to that alternative's share of the total attraction on offer. Let $w_s > 0$ denote the attraction of slot $s$ and $w_0 > 0$ the attraction of the always-available no-purchase alternative. The basic attraction model then specifies:
\begin{equation*}\label{eq:bam}
    \pi_s(\mathbf{x},\mathbf{p})
    =
    \frac{x_s\,w_s}{w_0 + \sum_{s \in S} x_s\,w_s},
    \qquad
    \pi_0(\mathbf{x},\mathbf{p})
    =
    \frac{w_0}{w_0 + \sum_{s \in S} x_s\,w_s}.
\end{equation*}
A logit model is a special case in which the attractions take the
exponential form $w_s = e^{v_s(p_s)}$, linked to the utilities through:
\begin{equation*}\label{eq:attraction_utility_map}
    w_s = e^{v_s(p_s)} \quad \Longleftrightarrow \quad v_s(p_s) = \ln w_s.
\end{equation*}
Substituting $w_s = e^{v_s(p_s)}$ and $w_0 = e^{v_0}$ gives:
\begin{equation*}\label{eq:bam_as_mnl}
    \pi_s(\mathbf{x},\mathbf{p})
    =
    \frac{x_s\,e^{v_s(p_s)}}{e^{v_0} + \sum_{s \in S} x_s\,e^{v_s(p_s)}},
\end{equation*}
which is exactly the softmax used in logit models over the offered slots together with the no-purchase option $v_0$, recovering the MNL probabilities where $v_{s} > 0$ denotes the attraction of slot $s$ and $v_{0} > 0$ the attraction of the always-available no-purchase alternative. Within AHD, BAM has seen limited adoption, and we mainly include it as a special case of the GAM employed by \cite{mackert2019} (see next paragraph) and to establish a conceptual link between logit models and attraction models. 

In the running example, suppose the retailer withdraws $A$ and $C$ and offers only $B$ (setting $x_A = x_C = 0$). The attractions $w_A$ and $w_C$ simply drop out of the denominator, and the probability they held is redistributed between the surviving slot $B$ and the no-purchase option in proportion to $w_B$ and $w_0$, similarly to MNL.

\paragraph{Generalized attraction model (GAM)}
\label{subsec:gam}
The generalized attraction model extends the basic attraction model by using both the offered slots and the slots that are \emph{not} offered. The mechanism attaches a ``dissatisfaction'' to the withheld slots: when a slot is withdrawn, the assortment is not simply reduced as in BAM; instead the withheld slot still contributes to the denominator, lowering the probabilities of the offered alternatives and pushing the no-purchase probability upwards. Introducing for each slot $s$ a dissatisfaction term $d_s \geq 0$, the GAM specifies:
\begin{equation*}\label{eq:gam}
    \pi_s(\mathbf{x},\mathbf{p})
    =
    \frac{x_s\,w_s}
         {w_0 + \sum_{s \in S}\big(x_s\,w_s + (1 - x_s)\,d_s\big)},
\end{equation*}
so that a slot which is \emph{not} offered ($x_s = 0$) still contributes $d_s$ to the denominator. The term $d_s$ captures the residual pull of the withheld slot. It represents demand that is \emph{partially lost} when slot $s$ is withdrawn, rather than fully redistributed to the offered alternatives as it would be under BAM. Setting every $d_s = 0$ recovers the basic attraction
model, and hence, through the exponential
mapping, any logit model. The GAM is thus the most general member of this family, with BAM and the logit models as successive special cases. \cite{mackert2019} models AHD choice through a GAM, homogeneous within delivery areas, with area-specific attraction values estimated from real-world UK e-grocer data. The central contribution is a mixed-integer-programming approximation of a request's opportunity cost, embedded in the cost term $\mathcal{C}(\mathbf{x},\mathbf{p})$, that jointly accounts for expected future demand-management decisions and their routing consequences and thereby captures the displacement cost of consumed capacity.

In the running example, withdrawing $A$ and $C$ (setting $x_A = x_C = 0$) now leaves their dissatisfaction terms $d_A$ and $d_C$ in the denominator. These residual terms absorb part of the demand that BAM would have pushed onto $\pi_B$, redirecting it toward $\pi_0$. As a result, some customers who wanted an early or late slot decline to substitute into the midday option and leave without booking. The recapture
into $\pi_B$ is therefore lower, and $\pi_0$ higher than under BAM, yielding a more conservative and empirically more plausible estimate of demand under a narrow assortment.

\vspace{-6pt}
\paragraph{Synthesis}
The review reveals that the customer choice models used in AHD are all instances of a random-utility formulation: 
\begin{equation*}\label{eq:utility_function}
    U_{is} = V_{is} + \varepsilon_{is},
    \qquad
    V_{is} = \boldsymbol{\beta}^{\top}\mathbf{x}_{is},
\end{equation*}
where $\mathbf{x}_{is}$ defines the attributes of the customer--slot pair (e.g., delivery fee $p_s$) and $\boldsymbol{\beta}$ is an estimated coefficient vector. Specific structures of $\boldsymbol{\beta}$ and the distribution of $\varepsilon_{is}$ recovers every model reviewed above, as illustrated below. 

First, the subscript on $\boldsymbol{\beta}$ selects the coefficient structure. A shared $\boldsymbol{\beta}$ has every customer weigh every attribute identically
(conditional logit); a slot-specific $\boldsymbol{\beta}_s$ lets each slot carry its own coefficients (MNL); and a customer-specific $\boldsymbol{\beta}_i \sim f(\cdot\mid\boldsymbol{\theta})$ treats the coefficients as random draws (mixed logit). Second, the distribution of $\varepsilon_{is}$ selects the structure of the residuals. I.i.d.\ Gumbel residuals yield the closed-form softmax at the cost of the IIA property; correlated residuals within nests (a GEV distribution) relax IIA within each nest, giving nested logit; and shrinking $\varepsilon_{is}$ toward zero makes choice deterministic, recovering the argmax rank-based rule. The discussed models are presented in Table~\ref{tab:choice_model_positioning}. The second and third column specify the structure of $V_{is} = \boldsymbol{\beta}^{\top}\mathbf{x}_{is}$ and the error distribution column specifies the form $\varepsilon_{is}$. 


\begin{table}[htbp]
\centering
\caption{Overview of choice models. Each model
corresponds to a combination of a coefficient structure, an attribute specification,
an error distribution, and a selection rule.}
\label{tab:choice_model_positioning}
\small
\begin{tabular}{@{}lllll@{}}
\toprule
Model & $\boldsymbol{\beta}$ structure & Attributes & Error distribution & Selection rule \\
\midrule
Conditional logit
  & Shared $\boldsymbol{\beta}$
  & $\mathbf{x}_s$ (slot-varying)
  & i.i.d.\ Gumbel
  & Softmax \\
MNL
  & $\boldsymbol{\beta}_s$
  & $\mathbf{x}_i$ (customer-varying)
  & i.i.d.\ Gumbel
  & Softmax \\
Mixed logit
  & $\boldsymbol{\beta}_i \sim f(\cdot\mid\boldsymbol{\theta})$
  & $\mathbf{x}_{is}$
  & i.i.d.\ Gumbel
  & Softmax integral \\
Nested logit
  & Any of the above, with nests
  & Hierarchical slots
  & GEV
  & Hierarchical softmax \\
Rank-based
  & Any
  & Any
  & $\varepsilon \to 0$
  & Argmax \\
BAM
  & Any (via $w_s$)
  & $w_s$
  & None
  & Attraction share \\
GAM
  & Any (via $w_s$)
  & $w_s$ and $d_s$
  & None
  & Attraction share \\
\bottomrule
\end{tabular}
\end{table}

\section{Experiments}\label{experiemtns}

In this section, we provide some results. In Section~\ref{expdesign} we provide the experimental design, and in Section~\ref{results} we provide the numerical results.

\subsection{Experimental Design}\label{expdesign}

We study the effect of the customer choice model on time-slot pricing in AHD, with the choice model as the only component that varies. There are $T = 6$ non-overlapping slots of two hours each, spanning 08:00 to 20:00, with midpoint hours $9, 11, \dots, 19$, and each slot $s$ has a serving cost $ \mathcal{C}(\mathbf{x},\mathbf{p})$. A customer is offered a subset $\mathbf{x}$ of the slots and either books one of them or does not buy. Each customer belongs to one of three segments, with weights $0.35, 0.30, 0.35$, that fix a preferred delivery hour $h_i \in \{10, 14, 18\}$, an eco-proneness $e_i \in \{0.6, 0, -0.4\}$, and a standardised income $\iota_i \in \{1, 0, -1\}$. These three attributes stand for the main observable reasons a customer prefers one slot over another: closeness to the hour they want, a taste for slots labelled as green, and a willingness to pay that moves with income. They also give the models something to disagree on, since the models differ in how much of this information they use.

\paragraph{Choice models.}
We build seven models on one shared utility, so that they differ only in which terms are active
and in the error structure. For customer $i$ and slot $s$,
\begin{equation*}\label{eq:sysutil}
    V_{is} = \alpha_s + \beta_p\,p_{is} + \beta_g\,g(h_i,s) + \beta_e\,e_i \ell_s
             + \beta^{\mathrm{inc}}_s\,\iota_i .
\end{equation*}
Here $\alpha_s$ is a slot constant, $p_{is}$ the fee, and $g(h_i,s) = \exp(-\tfrac{1}{2}((m_s - h_i)/\sigma_t)^2)$ a bell-shaped weight that is largest when the slot midpoint $m_s$ equals the customer's preferred hour $h_i$ and tails off as the two move apart, so it measures how well the slot time fits the customer. The term $e_i \ell_s$ rewards a match between an eco-conscious customer and a slot with a green label $\ell_s$, and $\iota_i$ is the income, with a slot-specific coefficient. The fee is a slot attribute and appears in every model. The models activate different parts of the utility: the conditional logit uses the shared terms $\beta_p, \beta_g,
\beta_e$; the MNL uses the slot constants $\alpha_s$, the fee, and the income term; and the hybrid uses all terms. The attraction models use $v_{is} = \exp(V_{is})$. Note that the BAM then coincides with the conditional logit, so we do not report it separately, and the GAM adds a term $\omega\,v_{iu}$ for each slot $u$ that is not offered, which sends the demand of a withheld slot toward not buying rather than toward the offered ones, with $\omega = 5$. The nested logit groups the slots into a morning, midday, and evening nest with dissimilarity $0.55$ and treats not buying as its own nest. The mixed logit gives each customer an individual price coefficient drawn from a normal distribution with mean $\beta_p$ and standard deviation $0.35$, and the rank-based model takes $\varepsilon \to 0$ and lets the customer pick the single highest-utility option, including not buying. Every model has a no-purchase option whose utility we fix at zero. We do not tune this value per model, so the models can differ in the share of customers who buy even before the fees have any effect.

\paragraph{Pricing.}
We set the delivery fees to maximise the expected profit of the arriving
customer under its own choice probabilities. For the logit and attraction models this maximum
has a closed form,
\begin{equation*}\label{eq:optprice}
    p^{*}_{s} =  \mathcal{C}(\mathbf{x},\mathbf{p}) - r - \tfrac{m}{\beta_p}, \qquad
    m = 1 + W\!\Big(\tfrac{G}{A\,e}\Big), \qquad
    G = \sum_{s \in \mathcal{H}} \exp\!\big(a_{is} + \beta_p ( \mathcal{C}(\mathbf{x},\mathbf{p}) - r)\big),
\end{equation*}
where $a_{is}$ is the utility without the fee, $r$ a fixed revenue per order,
$ \mathcal{C}(\mathbf{x},\mathbf{p})$ the cost of serving slot $s$, and $A$ the weight of the no-purchase option in the
denominator of the choice probabilities (one for the logit, or one plus the GAM term). See \citet{dong2009} and \citet{yang2016} for the optimality proof. The optimal fees share a common structure $-m/\beta_p$; the scalar $m$ follows from the first-order condition and is obtained with the Lambert $W$ function, the inverse of $x \mapsto x e^x$. For the nested, mixed, and rank-based models no such closed form formula exists, so we search for the profit-maximising fee vector with a numerical optimiser. Fees are limited to $[-15, 15]$.


\subsection{Numerical Results}\label{results}
We present the numerical results in three stages. First, we evaluate the misspecification of the customer choice model, i.e., the case in which the true model differs from the assumed model. Second, we show that even when the assumed model and the true model coincide, the sensitivity of its parameters lead to different outcomes. Third, we illustrate the impact the number of customer segments has on the expected profit.

All three experiments are analytical in the sense that the reported outcomes are based on the (closed-form) expected profit of an arriving customer under the optimal prices $\mathbf{p^*}$. Concretely, for a given set of time slots and customer segments, we compute the profit-maximising fees of the assumed model, evaluate the resulting choice probabilities under the true model, and the expectation $\sum_s \mathbb{P}(s)(r + p^*_s -  \mathcal{C}(\mathbf{x},\mathbf{p}))$. This isolates the choice-model because if, alternatively, analyses based on the realised profit confounds three effects: (i) the choice model, (ii) the arrival sequence of customers, and (iii) the routing costs.

\paragraph{Misspecification of the customer choice model.}
We study the impact of assuming the ``wrong" choice model when setting prices while a different, ``right", model is used for the customer's actual choices. In practice the ``right" model is never known. The purpose of this experiment is solely to show that a misspecified choice model changes the expected profit, i.e., customer choice models matter. 

\begin{figure}[!h]\centering
  \includegraphics[width=0.70\linewidth]{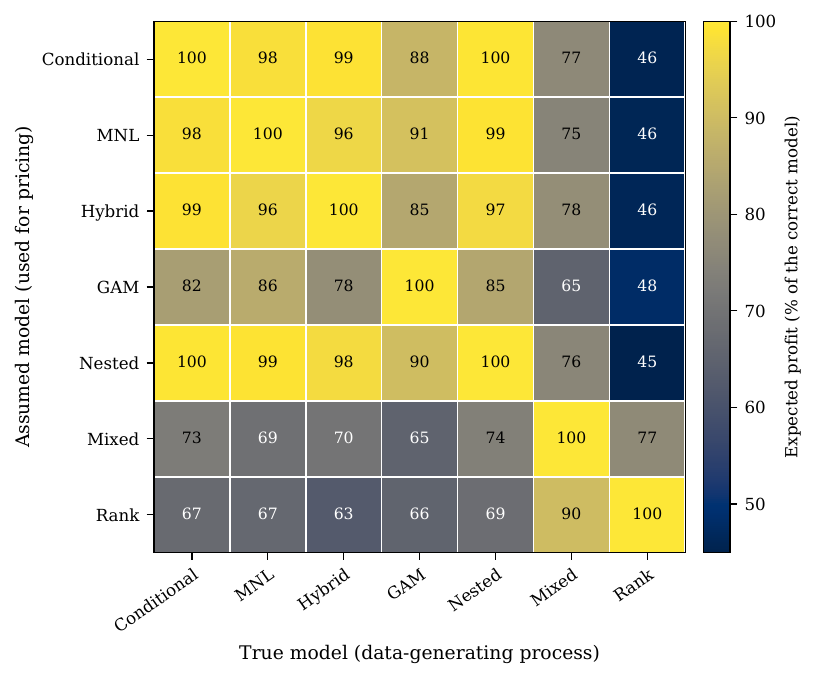}
  \caption{Expected profit of an arriving customer when the fees of an assumed model (rows) are
  charged while another model determines the choices (columns), as a percentage of the correctly
  specified model, averaged over the customer segments and the offer sets.}
  \label{fig:matrix}
\end{figure}

In Figure~\ref{fig:matrix} the rows denote the model whose profit-maximising fees are charged, and the columns are the model that determines the customers' choices. Each entry is the expected profit of an arriving customer as a percentage of the correctly specified model, averaged over the segments and the offer sets, so the diagonal is $100$ and every other entry is a loss. Among the four smooth logit models (conditional, MNL, hybrid, nested), pricing with the ``wrong" one loses at most $4\%$. The GAM differs more: pricing with the GAM under one of these models, or with one of them under the GAM, loses up to $22\%$. The largest losses are between these five models and the heterogeneous ones. When the rank-based model determines the choices, the five logit and attraction models keep only $45$ to $48\%$; when the mixed logit determines the choices, $65$ to $78\%$. In the other direction, using the mixed-logit model under one of the five models keeps $65$ to $74\%$, and the rank-based model keeps $63$ to $69\%$. This analysis leads us to the following fallacy. 

\begin{fallacy}
    \textbf{The known-model fallacy}
\end{fallacy}
\begin{fallacynote}
Because a correctly specified model prices optimally against itself, the diagonal of the matrix is 100 by construction. Hence, one is tempted to evaluate a pricing policy using the same choice model that generated it, implicitly assuming that the customers' true choice model is known.
The fallacy is avoided by not assuming the choice model is known. Rather than committing to a single estimated model, a candidate pricing policy should be tested against multiple plausible alternative choice models. Fees can then be chosen to hedge against any resulting uncertainty, for instance by maximising the worst-case or a weighted expected profit rather than the expected profit of any one model. A policy hedged in this way yields conclusions that remain robust even when the true model does not fully correspond to the assumed one.
\end{fallacynote}

\begin{figure}[!h]
    \centering
    \includegraphics[width=0.70\linewidth]{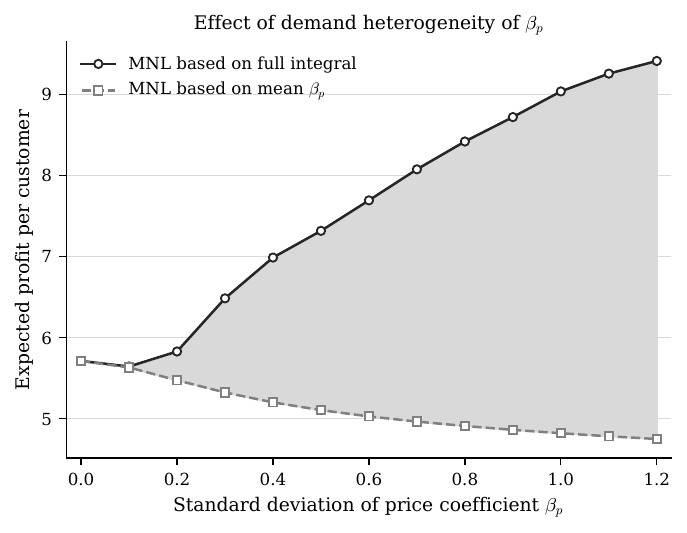}
    \caption{Expected profit per customer under the same MNL, evaluated either by integrating over the full distribution of the price coefficient (solid) or by substituting its mean (dashed) under different values of $\sigma$}.
    \label{fig:sensitivity_analysis}
\end{figure}

\paragraph{Lack of sensitivity analysis.}
Even when the assumed model and the true model coincide in its \textit{structure}, the true parameters are themselves estimated, and hence uncertain. To test the consequences of such uncertainty, we consider the assumed model and the true model to be the same (MNL) and only perturb the price sensitivity with $\varepsilon \sim \mathcal{N}(0, \sigma^2)$ resulting in $\beta_p := \beta_p + \varepsilon$, and show the expected profit as $\sigma$ grows. Figure~\ref{fig:sensitivity_analysis} shows the result of this experiment, where the expected prices are either derived based on the mean $\beta_p$ or based on the full integral, capturing the full extent of $\sigma$, i.e., a perfect information baseline. The expected profit per customer is highly sensitive to the assumed deviation of the price coefficient. In other words, MNL based only on the mean $\beta_p$ underestimates the expected profit when $\sigma > 0$, possibly leading to wrong insights in real-world applications for which $\sigma$ can fairly be assumed to be positive. The normally distributed price coefficient examined above is only one illustration; the same reasoning applies to any parameter whose true value is uncertain or heterogeneous. Nonetheless, most papers in the AHD literature report based on a single fitted parameter set and evaluate at that point alone, which leads us to the following fallacy.


\begin{fallacy}
    \textbf{The point-estimate fallacy}
\end{fallacy}
\begin{fallacynote}
A single fitted parameter set is only ever a best guess at a population that is, in truth, diverse. Just evaluating the mean of the believed model may be considered a good approximation, but it is not merely imprecise: because profit is a non-linear function of the underlying CCM parameters, the profit at the mean differs from the value averaged over the true parameter variation, in either direction. A policy evaluated at a point estimate may then report that heterogeneity hurts profitability when in fact it helps.
The remedy is to assess a policy across a distribution of parameter values rather than a single representative point.  This will yield insights into the population the retailer actually faces, rather than insights about a customer who may or may not exist in reality.
\end{fallacynote}

\begin{figure}[!h]
    \centering
    \includegraphics[width=0.72\linewidth]{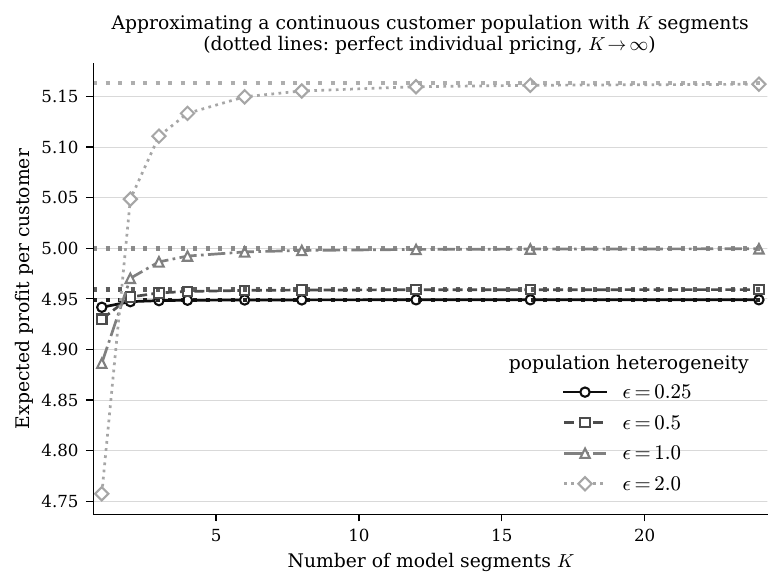}
    \caption{Expected profit per customer when a heterogeneous customer population is
approximated by $K$ discrete segments.}
    \label{fig:customer_segments}
\end{figure}

\paragraph{Too few customer segments.} Most papers in the AHD literature assume a single or a few customer segments. To quantify what this aggregation entails, we approximate a continuous customer population (in which each customer has an individual price sensitivity) by $K$ discrete segments. Customers are grouped into $K$ classes, each assigned a single $\beta_p$. We evaluate the expected profit as $K$ grows from one segment to the individual-level limit $K \to \infty$ (the dotted lines in Figure~\ref{fig:customer_segments}). The experiment is repeated at four levels of population heterogeneity $\epsilon$, which is the standard deviation of the true price coefficient across customers, ranging from a nearly homogeneous population
($\varepsilon = 0.25$) to a highly dispersed one ($\varepsilon = 2.0$), while keeping all other attributes and prices fixed.  From this figure, we conclude that the costs of using too few segments is controlled by $\varepsilon$. Note that the magnitude of the expected profit per customer is illustrative only. When the population is nearly homogeneous, single segment models are near-optimal, whereas using more customers segments improve performance when the population becomes more heterogeneous. In real-world applications, customer populations can assumed to be considerably heterogeneous, so that more segments are required than the single or few adopted by most state-of-the-art AHD methods.

\begin{fallacy}
    \textbf{The single-segment fallacy}
\end{fallacy}
\begin{fallacynote}
Estimating one parameter set for an entire delivery area or day can lead to significant approximation errors. A population that is assumed to be homogeneous in the model, but heterogeneous in reality, is priced based on one representative customer. The resulting optimal price for that average customer may in generally be optimal for exactly no one.
The fallacy is avoided by not presuming that one segment suffices. The population should be divided into as many segments that allow for a maximum acceptable approximation error. Otherwise, a policy built around a single (or a few) customers may look optimal when evaluating against these representative customers, yet leaving \href{https://youtu.be/IpwSXWq1wwU?t=59}{\underline{money on the table}} in reality.
\end{fallacynote}

\section{Conclusion}\label{conclusion}

To conclude, yes, customer choice models matter. Our three experiments show that the customer choice model is not an innocent modelling detail, but a decision that can change the reported profit and even
the direction of an effect. We identify three fallacies: (i) pricing with the same model as used in the evaluation can misspecify the expected profit when the true model is different (\textit{the known-model fallacy}), (ii) a single fitted parameter set misjudges a population that in truth has a standard deviation (\textit{the point-estimate fallacy}), and (iii) one or a few customer segments provide too little detail in heterogeneous populations (\textit{the single-segment fallacy}). In each case the remedy is the same: do not treat the customer choice model as known, but test a policy against a range of plausible models, parameters, and customer segments, and develop revenue management models that hedge against uncertainty. This asks for two things the field currently lacks, richer choice models that admit heterogeneity and relax the inconvenient independence of irrelevant alternatives property, and the disaggregate data to estimate them reliably.

Our insights are not specific to attended home delivery. The same four components, an offering and pricing policy, a customer choice model, and a cost approximation, underlie revenue management across last-mile logistics, and with them the same three fallacies. For instance, in out-of-home
delivery, choice models are used to select pickup locations, and policies are again derived and evaluated under the same assumed model. We have fallen victim to the fallacies ourselves, which is part of why we think they are worth being made explicit. We therefore hope the fallacies, and the call for better customer choice models, serve as a contribution beyond attended home delivery, whenever a customer choice model is embedded within an optimization problem.


\bibliography{main}

@article{yang2016,
  title={Choice-based demand management and vehicle routing in e-fulfillment},
  author={Yang, Xinan and Strauss, Arne K. and Currie, Christine S. M. and Eglese, Richard},
  journal={Transportation Science},
  volume={50},
  number={2},
  pages={473--488},
  year={2016},
  publisher={INFORMS}
}

@article{asdemir2009,
  title={Dynamic pricing of multiple home delivery options},
  author={Asdemir, Kursad and Jacob, Varghese S. and Krishnan, Ramayya},
  journal={European Journal of Operational Research},
  volume={196},
  number={1},
  pages={246--257},
  year={2009},
  publisher={Elsevier}
}

@article{campbell2006,
author = {Campbell, Ann and Savelsbergh, Martin},
year = {2006},
month = {08},
pages = {327-341},
title = {Incentive Schemes for Attended Home Delivery Services},
volume = {40},
journal = {Transportation Science},
}

@article{agatz2011,
  title={Time slot management in attended home delivery},
  author={Agatz, Niels and Campbell, Ann and Fleischmann, Moritz and Savelsbergh, Martin},
  journal={Transportation Science},
  volume={45},
  number={3},
  pages={435--449},
  year={2010},
  publisher={INFORMS}
}

@article{klein2019,
  title={Differentiated time slot pricing under routing considerations in attended home delivery},
  author={Klein, Robert and Neugebauer, Michael and Ratkovitch, Dimitri and Steinhardt, Claudius},
  journal={Transportation Science},
  volume={53},
  number={1},
  pages={236--255},
  year={2019},
  publisher={INFORMS}
}

@article{yang2017,
  title={An approximate dynamic programming approach to attended home delivery management},
  author={Yang, Xinan and Strauss, Arne K.},
  journal={European Journal of Operational Research},
  volume={263},
  number={3},
  pages={935--945},
  year={2017},
  publisher={Elsevier}
}

@article{cleophas2014,
  author    = {Cleophas, Catherine and Ehmke, Jan Fabian},
  title     = {When Are Deliveries Profitable? {Considering} Order Value and Transport Capacity in Demand Fulfillment for Last-Mile Deliveries in Metropolitan Areas},
  journal   = {Business \& Information Systems Engineering},
  volume    = {6},
  number    = {3},
  pages     = {153--163},
  year      = {2014},
  publisher = {Springer}
}

@article{ehmke2014,
  author    = {Ehmke, Jan Fabian and Campbell, Ann Melissa},
  title     = {Customer acceptance mechanisms for home deliveries in metropolitan areas},
  journal   = {European Journal of Operational Research},
  year      = {2014},
  volume    = {233},
  number    = {1},
  pages     = {193--207},
}

@article{mackert2019,
  author    = {Mackert, Jochen},
  title     = {Choice-based dynamic time slot management in attended home delivery},
  journal   = {Computers \& Industrial Engineering},
  year      = {2019},
  volume    = {129},
  pages     = {333--345},
}

@article{amorim2024,
  author    = {Amorim, Pedro and DeHoratius, Nicole and Eng-Larsson, Fredrik and Martins, Sara},
  title     = {Customer Preferences for Delivery Service Attributes in Attended Home Delivery},
  journal   = {Management Science},
  year      = {2024},
  volume    = {70},
  number    = {11},
  pages     = {7559--7578},
}

@article{Solomon1987,
  author    = {Marius M. Solomon},
  title     = {Algorithms for the Vehicle Routing and Scheduling Problems with Time Window Constraints},
  journal   = {Operations Research},
  volume     = {35},
  number     = {2},
  pages      = {254--265},
  year       = {1987},
}

@article{dong2009,
author = {Dong, Lingxiu and Kouvelis, Panos and Tian, Zhongjun},
year = {2009},
month = {04},
pages = {317-339},
title = {Dynamic Pricing and Inventory Control of Substitute Products},
volume = {11},
journal = {Manufacturing \& Service Operations Management},
}

@article{strauss2020,
  author  = {Strauss, Arne K. and G{\"u}lp{\i}nar, Nalan and Zheng, Yijun},
  title   = {Dynamic pricing of flexible time slots for attended home delivery},
  journal = {European Journal of Operational Research},
  volume  = {294},
  number  = {3},
  pages   = {1022--1041},
  year    = {2021},
  publisher = {Elsevier}
}

@article{abdolhamidi2024mixed,
  author  = {Abdolhamidi, Dorsa and Lurkin, Virginie},
  title   = {A tactical time slot management problem under mixed logit demand},
  journal = {OR Spectrum},
  year    = {2025},
  publisher = {Springer}
}

@book{Train2009,
  author    = {Kenneth E. Train},
  title     = {Discrete Choice Methods with Simulation},
  edition   = {2},
  publisher = {Cambridge University Press},
  address   = {Cambridge},
  year      = {2009},
}

@article{wamuth2023,
title = {Demand management for attended home delivery—A literature review},
journal = {European Journal of Operational Research},
volume = {311},
number = {3},
pages = {801-815},
year = {2023},
author = {Katrin Waßmuth and Charlotte Köhler and Niels Agatz and Moritz Fleischmann}
}
\bibliographystyle{tmlr}

\end{document}